\documentclass[usenatbib]{mn2e}
\usepackage{natbib}
\usepackage{psfig}
\title{The effects of similarity breaking on the intracluster medium} 
\author[E. J. Lloyd-Davies et al.]  
       {E. J. Lloyd-Davies$^{1,2}$,\thanks{E-mail: eld@astro.lsa.umich.edu}
        R. G. Bower$^3$, T. J. Ponman$^1$\\
        $^1$School of Physics and Astronomy, University of
        Birmingham, Edgbaston, Birmingham B15 2TT, UK\\
        $^2$Department of Astronomy, University of Michigan,              
                             Ann Arbor, MI 48109-1090\\
        $^3$Department of Physics, University of Durham,
        South Road, Durham DH1 3LE, UK} 
\date{Accepted 2002 ??.  
      Received 2001 ??; 
      in original form 2001 ??}

\pagerange{\pageref{firstpage}--\pageref{lastpage}}
\pubyear{2002}

\begin{document}

\maketitle

\label{firstpage}

\begin{abstract}
  We construct a family of simple analytical models of galaxy clusters at
  the present epoch and compare its predictions with observational data.
  We explore two processes that break the self-similarity of galaxy
  clusters: systematic variation in the dark matter halo concentration and
  energy injection into the intracluster gas, through their effects on the
  observed properties of galaxy clusters. Three observed relations between
  cluster properties and temperature are employed to constrain the model;
  mass, slope of gas density profile ($\beta$) and luminosity. The slope of
  the mass-temperature relation is found to be reproduced by our model when
  the observed variation in concentration is included, raising the
  logarithmic slope from the self-similar prediction of 1.5, to that of the
  observed relation, $\sim 2$. Heating of the intracluster gas is observed
  to have little effect on the mass-temperature relation. The mean trend in
  the $\beta$-temperature relation is reproduced by energy injection in the
  range 0.5-0.75 keV per particle, while concentration variation is found
  to have only a small effect on this relation. Excess energies calculated
  for individual systems from the $\beta$-temperature relation suggest that
  the lowest mass systems may have excess energies that are biased to lower
  values by selection effects. The observed properties of the
  luminosity-temperature relation are reproduced by the combined effects of
  excess energy and a trend in the dark matter concentration. At high
  masses the observed variation in dark matter concentration results a
  logarithmic slope of $\sim 2.7$ compared to recent observations in the
  range 2.6-2.9, whilst the observed steepening of the relation in galaxy
  groups is predicted by the model when heating in the range 0.5-0.75 keV
  per particle is included.  Hence a combination of energy injection and
  dark matter concentration variation appears able to account for the mean
  trends in all the observed relations. Scatter in the energy injection and
  concentration may account for a large proportion of the scatter in the
  observed relations.
\end{abstract}


\section{Introduction}

In the hierarchical clustering model for the formation of structure in the
universe, small scale perturbations collapse into virialized objects and
then cluster together to form successively larger virialized structures.
Simple physical models of gravitational collapse along with shock heating
of the gas would suggest that these virialized structures should be
approximately scaled versions of one another \citep{navarro95a}. In the
absence of other physical process such as heating or cooling, the gas and
dark matter halos will be almost self-similar. It is possible for energy to
be transferred between the dark matter and the gas during mergers
\citep{pearce94a} which is believed to be responsible for the cores
observed in cluster gas density profiles. However this is not usually
assumed to act in a way which would break self-similarity.

This expected self-similarity has already been observed to be broken in
several respects. The surface brightness profiles of galaxy clusters have
been observed to flatten in low mass systems \citep{ponman98a,helsdon00a}
and their gas density, entropy and energy have been shown not to be
self-similar \citep{lloyd-davies00a}. The luminosity-temperature relation
for galaxy clusters is also observed deviate from its expected self-similar
behaviour \citep{edge91a,markevitch98a,helsdon00b}. These effects are
usually attributed to energy injection into the intracluster medium (ICM)
from galaxy winds \citep{lloyd-davies00a}.  However there is some
controversy over whether galaxy winds can provide the energy necessary to
heat the ICM \citep{valageas99a,wu99a,bower00a}. An alternative often
promulgated is that AGN are responsible for the injected energy. However
while the amount of energy available from AGN is very large, it is not
entirely clear how they could transfer it into the ICM. AGN have emitted
large amounts of energy in the form of electromagnetic radiation,
especially in the ultra-violet, but this will not heat gas to the required
temperatures. Mechanical energy transfer through AGN jets seems the only
possible option at present, but it is not clear how much energy is
available in this form or whether it can be transferred efficiently into the
ICM. It should be noted that preliminary Chandra results \citep{fabian01a}
do not suggest that widespread heating of the ICM by AGN is occurring at low
redshift.

The amount of energy injection inferred from the observed breaking of
similarity is also controversial. \citet{wu99a} suggest a value of 1.8-3.0
keV per particle is needed to explain the steepening of the
luminosity-temperature relation from the self-similar prediction of $\sim
T^2$ to the observed value of $\sim T^3$. However \citet{lloyd-davies00a}
have measured a mean excess energy of 0.44 keV per particle in a sample of
20 galaxy clusters and groups using analytical models fitted to X-ray
spectral images. There would also seem to be a theoretical argument against
energy injection of much greater than 1 keV per particle since it is
difficult to see how galaxy groups with mean temperatures less than 1 keV
could have X-ray haloes, whereas a number of such systems are known to
exist \citep{ponman96a,helsdon00a}. It therefore seems unlikely that energy
injection into the the ICM can be the explanation for the steepness of the
\emph{cluster} luminosity-temperature relation or that any energy injection
can greatly exceed 1 keV per particle.

Breaking of simple self-similarity in the dark matter constituents of
galaxy clusters has also been observed. The dark matter concentration of
galaxy clusters systematically decreases with increasing cluster mass
\citep{wu00a,sato00a,lloyd-davies00b}. This effect has been observed for
some years in numerical simulations \citep{navarro97a,moore99a}. It is
generally interpreted as due systematic variations in the formation time
with system mass. This is predicted by the hierarchical clustering model of
structure formation, with low mass systems collapsing first and then
clustering together to form progressively larger structures. The exact
process by which this affects the dark matter concentration is a matter of
some uncertainty, but an interesting hypothesis is that halos all form with
the same concentration, and subsequent accretion increases the virial radius
and therefore the concentration \citep{salvador-sole98a}. Therefore halos
that form at high redshift and therefore have had a long time to accrete
material will have the highest concentrations. As the gravitational
potential affects the temperature structure of the ICM, variation of the
dark matter concentration would be expected to have an impact on the
observed properties of the ICM.

We therefore have two processes which are observed to break self-similarity
in galaxy clusters. Heating of the ICM which results in flattening of the
gas density profile, and systematic variation in the dark matter halo
concentration. In order to put constraints on these processes one useful
approach is to construct models of galaxy clusters and then compare the
predictions of the models with observed properties of clusters. A similar
approach has previously been used by a number of authors
\citep{cavaliere97a,wu98a,wu99a,cavaliere99a,balogh99a,tozzi00a,bower00a} to
explore the effects of energy injection. However these models have in
general been driven from a theoretical perspective and have also tended to
concentrate their efforts on predicting the evolution of cluster properties
with redshift. Our aim is to construct a model of galaxy clusters, base on
empirical results wherever possible, which will predict their properties at
zero redshift. We will then use this model to investigate in detail how
processes that break self-similarity, such as injection of energy into the
ICM, affect the systems' observed properties. Comparison of the model
predictions with observed data should then allow constraints to be placed
on the various process involved in breaking self-similarity.

\section{Cluster Model}

In order to construct useful models of galaxy clusters to be compared with
observations, a number of simplifying assumptions need to be made. In this
case we will use only one-dimensional spherically symmetric models
containing only dark matter and hot gas in hydrostatic equilibrium. While
these approximations are crude, it should be noted that most analyses of
observations of galaxy clusters involve the assumption of spherical
symmetry. The approximation is fairly good for relaxed systems but it
should be born in mind when considering the results that it is not a very
good approximation to morphologically disturbed systems. In our model we
make no distinction between dark matter and galaxies. However to a first
approximation galaxies can be considered to be collisionless and while
dynamical friction may be have some effect on the galaxies in low mass
systems, it is unlikely that this will significantly affect the
gravitational potential given that the galaxies make up only a small fraction
of the mass of the system. Our approach is not to model the evolution of
galaxy clusters to the state we observe them at present, but to model a
variety of possible end points of cluster evolution and then compare them
against observed clusters.

The primary parameter in the cluster model is the cluster mass. There are a
number of possible ways the cluster mass could be defined. In general the
mass within the region around the cluster centre that is in virial
equilibrium is considered. The overdensity of this region with respect to
the critical density, $\rho_{crit}$, can be calculated for the collapse of
a spherical top-hat density perturbation \citep{peebles80a}. For a critical
density universe this overdensity is $18 \pi^{2}$ ($\sim 178$) but can be
significantly smaller for other cosmologies \citep{bryan98a}. An overdensity
value of 200 is often used to define the outer boundary of a system as this
is smaller than the virial radius for all reasonable cosmologies. In order
for our results to be easily comparable with observations this overdensity
was used. This results in a relationship between the mass, $M_{200}$, and
the radius, $R_{200}$, of
\begin{equation}
R_{200}=\left(\frac{3 M_{200}}{800 \pi \rho_{crit}}\right)^{\frac{1}{3}}, 
\end{equation}
which depends only on the critical density, $\rho_{crit}$. Throughout the
modeling we adopt $H_{0}$=50~km s$^{-1}$ Mpc$^{-1}$ and $q_{0}$ = 0.5.

\subsection{ICM density profile}
In our model the gas density of the ICM is represented by the usual
parameterization of the form,
\begin{equation}
\rho(r)=\rho(0)\left[1+\left(\frac{r}{r_{c}}\right)^{2}\right]^{-\frac{3}{2}\beta},
\label{eq:beta}
\end{equation}
where $r_{c}$ is the core radius, $\beta$ is the density index and
$\rho(0)$ is the density normalization. \citet{jones84a} have found this to
be a good representation of the structure of the ICM in galaxy clusters.
\citet{vikhlinin99a}, using the best quality ROSAT data, agree with this,
although they find marginal evidence for steepening of the surface
brightness profile at large radius. The fiducial value for the $\beta$
parameter in our model was the canonical value of $\frac{2}{3}$. A fiducial
gas fraction within $R_{200}$ of 0.2, as observed by
\citet{lloyd-davies00b}, was used to set the normalisation of the gas
density profile.

To define the core radius of the gas density profile, $r_{c}$, we use the
value of 7 percent of $R_{200}$ found by \citet{lloyd-davies00b} for
systems where a cooling flow does not obscure the core. Selecting a value
for the normalization of the gas density profile, $\rho(0)$, is not a
trivial problem. One possibility is to normalize the profile such that the
gas mass within $R_{200}$ is $f_{gas}M_{200}$ where $f_{gas}$ is the gas
fraction of the cluster and a parameter to be specified. However this
method may be unphysical as it does not allow gas to be pushed outside
$R_{200}$ when the specific energy of the gas is raised by a large amount.
The normalisation of the gas density is therefore a boundary condition for
the problem which needs careful consideration.

\subsection{Dark matter density profile and gravitational potential}
In our model no distinction between dark matter and stars is made, since
both are expected to be collisionless on a large scale and neither will
contribute significant emission to the X-ray data we will be comparing the
model against. The dark matter distribution in our model is represented by a
profile derived from numerical simulations \citep{navarro95a} of the form,
\begin{equation}
\rho_{DM}(r)=\bar{\rho}_{DM}\left[x(1+x)^{2}\right]^{-1},
\end{equation}
where $x=r/r_{s}$ and $r_{s}$ is a scale radius. It should be noted that it
has recently been suggested \citep{moore99a} that higher resolution
simulations produce dark matter profiles with steeper central cusps, with
asymptotic slopes of about 1.5 rather than 1 in the profile of
\citep{navarro95a}. However this effect is only important at small radii
and as many observational results are derived using the profile of
\citep{navarro95a} it is better suited to our purposes. It is possible to
parameterize the concentration of the dark matter profile by the
concentration parameter, $c=r_{200}/r_{s}$\citep{navarro97a}. This
concentration is seen to be anti-correlated with cluster mass in numerical
simulations \citep{navarro97a} and this has been observed using X-ray data
\citep{wu00a,sato00a,lloyd-davies00b}. It is possible for us to take
account of this variation in concentration using an analytical relation,
for instance the relation
\begin{equation}
c=34.9\left(\frac{M_{200}}{10^{13}M_{\odot}}\right)^{-0.51},
\label{beq:concentration}
\end{equation}
is measured by \citet{lloyd-davies00b} for a sample of 20 galaxy clusters
and groups. It should be noted that this relation is a rough
characterization of a trend that shows considerable scatter. It is also
possible to use the results of numerical simulations to define the dark
halo concentration. The simulations of \citet{navarro97a} predict
concentration variation for a $\Lambda$CDM cosmology that can be approximated
by the relation
\begin{equation}
c=13.2\left(\frac{M_{200}}{10^{13}M_{\odot}}\right)^{-0.20},
\label{beq:concentration2}
\end{equation}
over the mass range of galaxy clusters and groups. This predicted relation
is considerably flatter than the those observed by
\citet{wu00a,sato00a,lloyd-davies00b}.  We will investigate the use of both
relations. The normalization of the of the dark matter profile,
$\bar{\rho_{DM}}$, is set so that the cluster contains a dark matter mass
of $(1-f_{gas})M_{200}$ within $R_{200}$.

With the gas and dark matter profiles defined it is then possible to
calculate the cluster temperature profile assuming hydrostatic equilibrium.
For hydrostatic equilibrium and spherical symmetry, the equation
\begin{equation}
M(r)=-\frac{T(r)r}{G\mu}\left[\frac{dln\rho}{dlnr}+\frac{dlnT}{dlnr}\right]
\end{equation}
is satisfied \citep{fabricant84a}. Given the mass and gas density profiles
of a cluster the temperature profile is defined if the temperature is
specified at some point. This introduces a second boundary condition to
the problem that we must consider.

\subsection{Effects of ICM cooling}

In our model we do not explicitly model the effects of radiative cooling on
the ICM for several good reasons. The amount of cooling in a particular
cluster is will be related to the amount of time it has been left
undisturbed by a major merger. This is an essentially random factor which
varies from system to system, depending on its particular history.  In
contrast, our model aims to represent the present state of clusters, and to
avoid the complications and uncertainties associated with cluster
evolution. Secondly while the physics associated with radiative losses
from ion-electron interactions in the ICM is well understood, the
macroscopic effects of this cooling are not. The cooling is thought to
result in a highly multiphase structure in the ICM but insufficient details
are known to accurately model it. Thirdly, current observations do not
provide enough information to provide a secure empirically based model of
cooling in clusters.

We therefore adopt an approach of modeling the clusters without any cooling
included and then attempting to take account of this when comparisons are
made with observational data. In many cases it is possible to use
observational data that has been corrected for the effects of cooling.
This should allow a reasonably unbiased comparison of the model predictions
with the data. In some cases, especially for galaxy groups cooling
corrected data is not available. In these cases we will attempt to estimate
what the likely effects of this are on our results using empirical data on
correction for cooling. While this is far from ideal we believe that
attempting to model the effects of cooling would introduce at least a
similar amount of uncertainty into our results, given our present state of
knowledge of cooling flows.

\subsection{Boundary conditions}

For a cluster that has the specific energy of the gas raised above the
default value by flattening its gas density profile, there are two boundary
conditions that must be specified. A normalisation for the gas density
profile and a normalisation for the temperature profile. The default
clusters need only one boundary condition, the temperature profile
normalization, as the gas density normalisation is set by the gas fraction
within the virial radius. However for the clusters with raise specific
energy if the model is to allow the gas fraction to vary another constraint
must be found. The most physically justifiable places to set the boundary
conditions are at the centre of the systems and at the shock radius.
However the position of the shock will be dependent on the amount of time
that has elapsed since the system formed and also on the extent to which
the infalling gas has been previously heated. Simulations suggest that the
shock radius should occur at $1-1.5 R_{vir}$ \citep{knight97a,tozzi00a} and
that preheating can result in the shock propagating out as far as $2.5
R_{vir}$ \citet{tozzi00a}, since higher entropy gas has a higher sound
speed.

One possible way of setting the density boundary condition is to use the
observation of \citet{lloyd-davies00b} that the gas density extrapolated to
$R_{200}$ appears to converge to an approximately constant value for
systems over a wide range of system masses. It should be noted that this
result relies on extrapolating models fitted to data within a radius
smaller than $R_{200}$ and is not a direct measurement of the density at
$R_{200}$. The gas density at $R_{200}$ in the raised specific energy case
can therefore be fixed at the density in the default case. This approach
has the advantage of being extremely simple and has a least some
justification from observations. We will therefore use this condition for
the main results in the paper but will investigate the effect of
alternative assumptions in Section \ref{sec:boundary}.

A constraint is also needed to set the temperature boundary condition for
the model. Unfortunately observed temperature profiles of galaxy clusters
are much less well constrained than gas density profiles and therefore
there are not much in the way of observational constraints on the
temperature at $R_{200}$. Therefore it is necessary to resort to
theoretical constraints on the temperature profile. A constraint
\citet{bower00a} derived from the numerical simulations of \citet{eke98a}
and \citet{frenk00a} is that the temperature at the virial radius is
approximately $0.5 T_{vir}$, where $T_{vir}$ is defined by the equation,
\begin{equation}
T_{vir}=\frac{\mu G M_{200}}{2 k R_{200}}.
\label{eq:tvir}
\end{equation}
The mean mass per particle $\mu$, is taken to be 0.6 amu.

\begin{figure*}
\begin{minipage}{140mm}
\centering{
\vbox{\psfig{figure=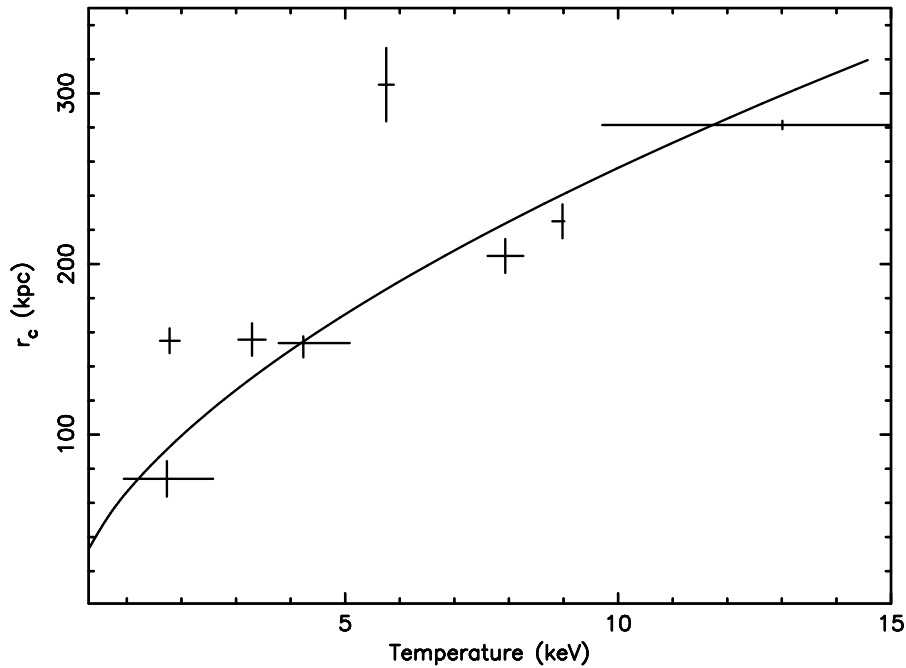}}\par
}
\caption{Core radius plotted against emission weighted temperature for 
  the 8 galaxy clusters and groups in the sample of
  \protect\citet{lloyd-davies00b} with reliable core radii measurements.
  The solid line shows the relation predicted by the model between core
  radius and emission weighted temperature when the core radius is a
  constant fraction, 7 percent, of $R_{200}$.}
\label{bfig:plot0}
\end{minipage}
\end{figure*}

\subsection{Energy calculation}

Observational studies \citep{lloyd-davies00b,helsdon00a} suggest that the
ICM density profiles in low mass systems are flattened compared to the
profiles of high mass systems and this is observed as a reduction of the
$\beta$ parameter in low mass systems. There is little evidence to suggest,
although it cannot be ruled out at present, that there is any significant
effect on the core radius of the gas density profile. Figure
\ref{bfig:plot0} shows the core radii of the 8 galaxy cluster and groups in
the sample of \citet{lloyd-davies00b} with reliable core radii
measurements, plotted against emission weighted temperature.  The solid
line shows the relation predicted by the model between core radius and
emission weighted temperature when the core radius is a constant fraction,
7 percent, of $R_{200}$. It can be seen that while the constraints of the
data are not very strong, a model where the core radius is a fixed fraction
of $R_{200}$ appears to be quite consistent with the data. Since
observational evidence strongly favours variation in the $\beta$ parameter
rather than the core radius \citep{lloyd-davies00b,helsdon00a,mohr99a}, this
approach therefore seems the most justifiable to take in modeling
variation in the gas specific energy.

Our model therefore allows the gas specific energy to be raised by reducing
the value of the $\beta$ parameter and so flattening the gas density
profile. In order to quantify difference in gas specific energy from the
default case, a prescription is needed for how the total energy of the ICM
is to be measured. It is possible given the temperature and density
structure of the ICM, and also the dark matter distribution, to calculate
the thermal and gravitational potential of the ICM. However it has been
pointed out by \citet{lloyd-davies00a} that care must be taken in selecting
the regions over which to calculate the energy as if the gas distribution
is changed the energy of the ICM within a fixed radius will not be
comparing the same mass of gas. To calculate the difference in the energy
of the ICM between raised specific energy and default models we therefore
calculate the difference between the total energy of the gas within
$R_{200}$ in the raised specific energy model and the total energy of the
same mass of gas in the default model, which will be contained within a
smaller radius.  Altering the value of $\beta$ affects both the
gravitational potential and thermal energy of the gas, since the gas
temperature profile will be modified if the gas density profile is
changed. Using this prescription it is possible to alter the value of
$\beta$ until the desired difference in specific energy is achieved. This
approach is similar to that used in the model of \citet{bower00a}.

\section{Observational data}
\label{sec:obs}
In order to constrain the parameters of our cluster models observational
data is needed to compare with the models predictions. There are a number
of observed relations which the model should reproduce in order to be an
accurate representation of galaxy clusters. For instance the mass
temperature relation is a fundamental relation which the model will need to
reproduce.  The mass-temperature relation of \citet{lloyd-davies00b} for 20
systems over a large mass range was used to compare with the model
predictions. The masses were derived from analytical models fitted to X-ray
spectral images and the temperatures are cooling flow corrected. This is
advantageous as our model does not model the effects of radiative cooling
in cluster cores.

An observed $\beta$-temperature relation to compare against our model
predictions were obtained from \citet{lloyd-davies00b} (20 systems) and
\citet{helsdon00a} (6 systems). \citet{lloyd-davies00b} fitted spherically
symmetric gas density profiles to X-ray spectral images of galaxy clusters
and groups while parameterizing either the gas temperature profile or dark
matter distribution by some functional form.  \citet{helsdon00a} fitted
surface brightness profiles to X-ray images of galaxy groups while allowing
ellipticity. Their $\beta$ values are therefore only strictly comparable to
the gas density profiles of \citet{lloyd-davies00b} on the assumption of
isothermality. Also the model fitting of \citet{lloyd-davies00b} assumes
spherical symmetry whereas \citet{helsdon00a} do not, and the temperatures
of \citet{lloyd-davies00b} are cooling flow corrected whereas those of
\citet{helsdon00a} are not. Despite these differences the results obtained
for these to studies are remarkably similar. 

The luminosity-temperature relation data we use to compare with our model
predictions is taken from \citet{markevitch98a} and \citet{arnaud99a} for
galaxy clusters and \citet{helsdon00b} for galaxy groups. The data of
\citet{markevitch98a} have been corrected for the effects of cooling and
the sample of \citet{arnaud99a} was chosen not to contain any systems
significantly affected by cooling. However the data of \citet{helsdon00b}
has not been corrected for the effects of cooling and there is not much
other data available for these low temperature systems. Therefore when
comparing them with the model predictions we will attempt to quantify the
likely effects on our results of the galaxy groups being uncorrected. It
should also be noted that the radii from within which luminosity data are
extracted are not consistent. This is more of a problem for the low mass
systems which have flatter surface brightness profiles. In these systems
the luminosity has generally been derived within approximately 0.3
$R_{200}$ whereas for more massive system the luminosity derived within
some larger radius.

\section{Results}
A fundamental parameter in all the relations we are interested in is the
gas temperature. We therefore consider how best to extract a gas
temperature from our model and then go on to compare the predictions of the
model with the relations listed in Section \ref{sec:obs}. Finally we will
investigate how the boundary conditions affect the results in Section
\ref{sec:boundary}.

\subsection{Temperature derivation}

The effects of raising the specific energy of the gas or increasing the
dark matter concentration on the gas temperature are quite similar. In both
cases the central temperature and temperature gradient are increased. The
effects of increasing the dark matter concentration are more centrally
concentrated however, with most of the temperature increase occurring near
the core of the system. In contrast raising the specific energy of the gas
raises the gas temperature out to quite large radii. In terms of the effect
on the emission-weighted temperature increasing the dark matter
concentration is fairly straight forward, since it has no effect on the gas
density profile and hence the emission-weighting is not significantly
altered. Increasing the dark matter concentration therefore results in an
increase in the emission-weighted temperature. In the case of raising the
specific energy of the gas the flatter gas density profiles result in the
emission being weighted more towards larger radii where the gas temperature
is lower. The effect on the emission-weighted temperature is therefore not
simple and in general depends on the radius within which the
emission-weighted temperature is calculated. For temperatures calculated
within $R_{200}$ raising the specific energy of the gas has little or no
effect whereas temperatures calculated within 0.3 $R_{200}$ rise as the
specific energy of the gas is raised.

Perhaps the most natural way to derive gas temperatures from our model is
to calculate emission-weighted temperatures within $R_{200}$ since the the
observed temperatures that we will be using will be weighted in this way.
However it should be noted that most actual observations of galaxy clusters
do not extend to this radius. In the case of the low mass systems that we
are particularly interested in, 0.3 $R_{200}$ is more representative of the
radii within which emission-weighted temperatures are derived. Even in high
mass systems the X-ray data rarely extend out to anywhere near $R_{200}$.
We therefore use 0.3 $R_{200}$ as the radius within which we extract
emission-weighted temperatures. It should be noted that in general the
radius to which X-ray data are available increases with increasing system
temperature and this may lead to some bias in the observed data. As
previously noted the model does not contain any cooling whereas in many
observed systems cooling has some effect on their emission-weighted
temperatures. In some cases it is possible to obtain data corrected for the
effects of cooling and where possible we make use of it.

\subsection{Mass-temperature relation}

\begin{figure*}
\begin{minipage}{140mm}
\centering{
\vbox{\psfig{figure=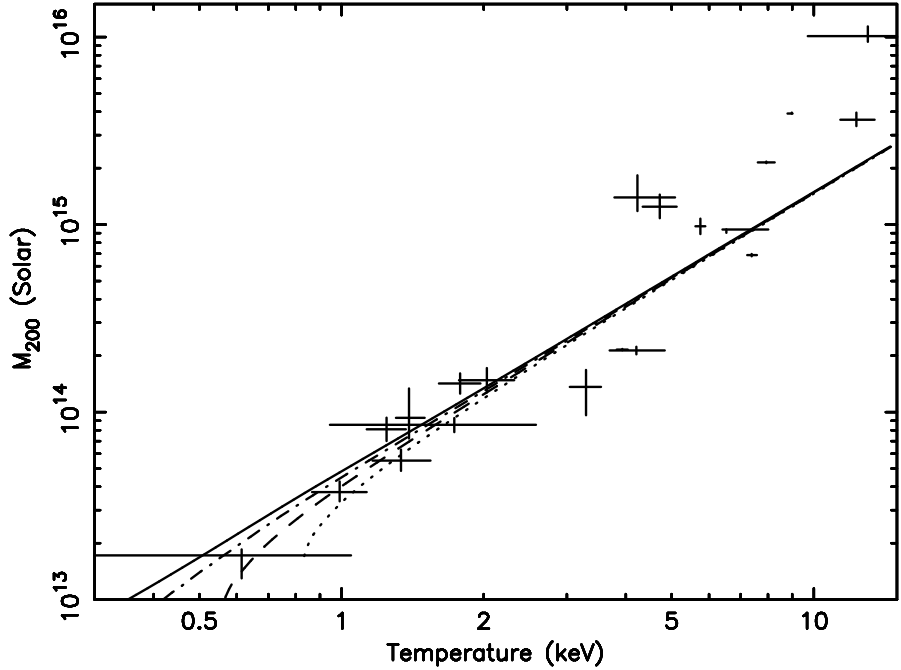}}\par
}
\caption{Total mass within $R_{200}$ against emission weighted temperature
  within 0.3 $R_{200}$ for simulated relations with the specific gas energy
  raised various amounts; 0.0 keV per particle (solid); 0.25 keV per
  particle (dot-dashed); 0.5 keV per particle (dashed); 0.75 keV per
  particle (dotted). A constant concentration parameter of 10 is used for
  all the relations. Observed data (crosses) are taken from the sample of
  \protect\citet{lloyd-davies00b}.}
\label{bfig:plot7}
\end{minipage}
\end{figure*}

An important relation which will constrain our models is the mass
temperature relation. Raising the specific energy of the ICM can affect its
temperature, and as the flattening of the density profile can push gas
outside the virial radius, the mass within $R_{200}$ can also be affected
to a certain extent. Self-similar theory predicts that the mass should be
proportional to $T^{\frac{3}{2}}$ and this is generally seen in numerical
simulations \citep{navarro95a}.  Figure \ref{bfig:plot7} shows the total
mass within $R_{200}$ plotted against emission weighted temperature within
0.3 $R_{200}$ for a range of cluster masses. The systems all have a
constant concentration parameter of 10. To compare our simulated
mass-temperature relations to observations, we use data from
\citet{lloyd-davies00b} for their sample of 20 galaxy clusters and groups.
The solid line shows the mass-temperature relation for the default model.
As expected from self-similar scaling the relation is a powerlaw with an
index of 1.5.  The other lines show the relations for systems where the
specific energy has been raised by 0.25 keV per particle (dot-dashed), 0.5
keV per particle (dashed) and 0.75 keV per particle (dotted).  It can be
seen that the amount of deviation from the self-similar mass-temperature
relation is not large, even when the specific energy of the gas is raised
by a large amount. It should be noted that in the 0.75 keV per particle
(dotted) case line terminates before the lowest mass is reached as beta is
flattened to 0 at this point.  It therefore seems that large deviations
from the mass-temperature relation due to heating are not possible, at
least with our model.

\begin{figure*}
\begin{minipage}{140mm}
\centering{
\vbox{\psfig{figure=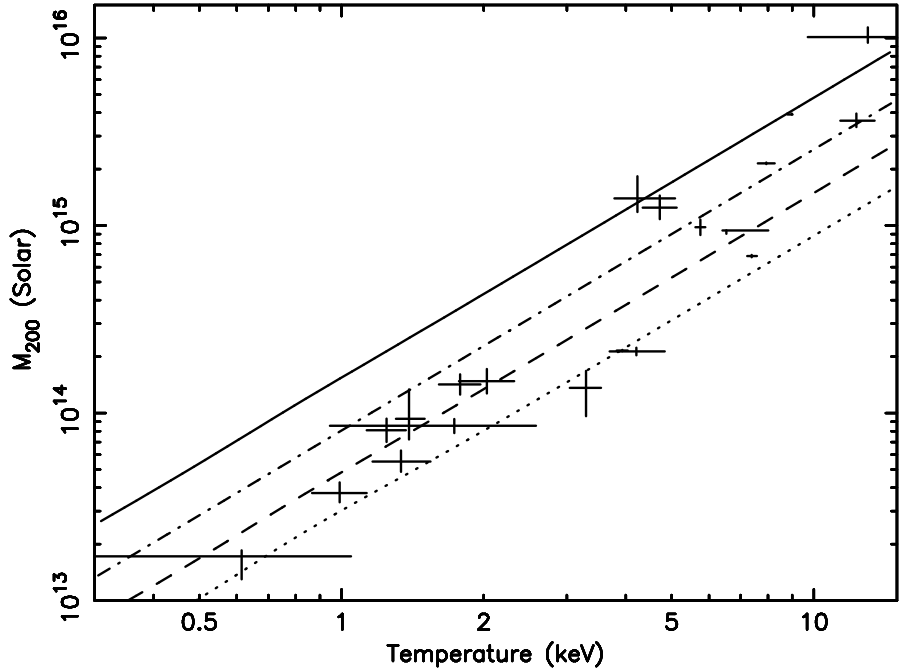}}\par
}
\caption{Total mass within $R_{200}$ against emission weighted temperature
  within 0.3 $R_{200}$ for simulated relations with various dark matter
  concentrations; c=2 (solid); c=5 (dot-dashed); c=10 (dashed); c=20
  (dotted). The excess energy for all the relations is zero. Observed data
  (crosses) are taken from the sample of\protect\citet{lloyd-davies00b}.}
\label{bfig:plot8}
\end{minipage}
\end{figure*}

The other process that might affect the mass temperature relation is the
variation in dark matter concentration. The concentration will not affect
the mass within $R_{200}$ but it will have some effect on the temperature.
Figure \ref{bfig:plot8} shows the total mass within $R_{200}$ plotted
against emission weighted temperature within 0.3 $R_{200}$ for several
values of the concentration parameter. It can be seen that as the
concentration increases from c=2 (solid line) to c=20 (dotted line) the
normalisation, defined as the mass at 1 keV, decreases from
$1.5\times10^{14}M_{\odot}$ to $3\times10^{13}M_{\odot}$. This can be
compared with other relations from observations and theory. The NFW
relation obtained from numerical simulations has a normalisation of
$8.7\times10^{13}M_{\odot}$ \citep{navarro95a} which is in between our two
extreme relations. \citet{lloyd-davies00b} have fitted a $T^{\frac{3}{2}}$
powerlaw to this sample of 20 galaxy clusters and groups and obtained a
normalisation of $9.7\pm1.4\times10^{13}M_{\odot}$ which is also bracketed
by our simulated relations. However this observed normalisation is for all
galaxy systems.  \citet{lloyd-davies00b} found that for systems with
temperatures below 4 keV the normalisation was
$3.1\pm0.4\times10^{13}M_{\odot}$ which is comparable with our high
concentration model relations. This can be understood in terms of the
variation in the concentration parameter with system mass. We have seen in
Figure \ref{bfig:plot8} that the dark matter concentration has a direct
effect on the normalisation of the mass-temperature relation. Since lower
mass systems are more concentrated, on average, their mass-temperature
relation would be expected to have a lower normalisation. Whereas rich
clusters of galaxies with concentration parameters of around 5 would be
expected to have a much higher normalisation to their mass-temperature
relation, which is what \citet{lloyd-davies00b} observe. The actual mean
mass-temperature relation should therefore be steeper than the simple
$T^{\frac{3}{2}}$ relation expected from self-similar scaling, due to the
increase in concentration in lower mass systems. This effect is indeed
observed by \citet{lloyd-davies00b} who observe the relation to have a
logarithmic slope of $1.96\pm0.21$.

\begin{figure*}
\begin{minipage}{140mm}
\centering{
\vbox{\psfig{figure=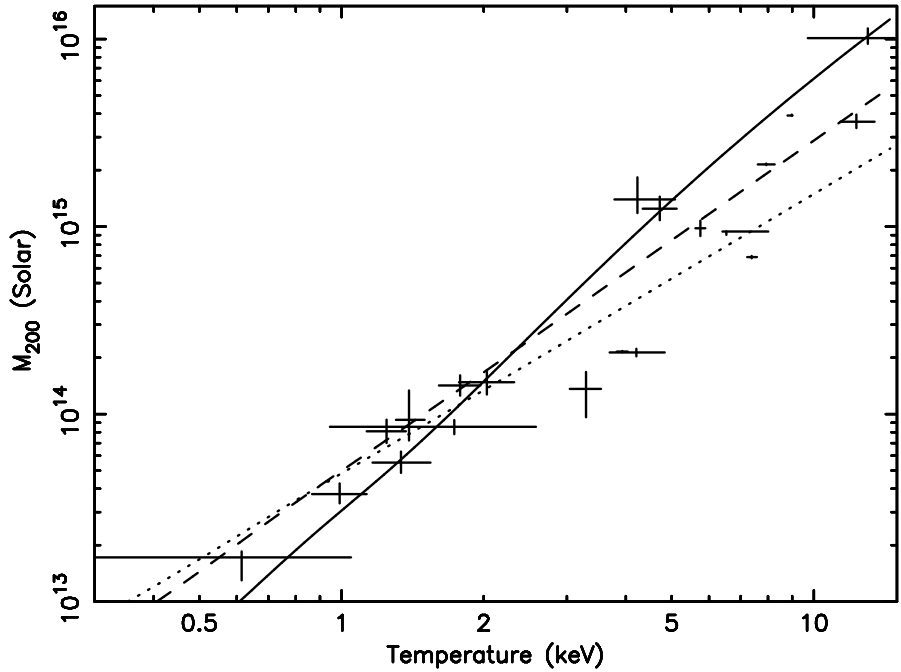}}\par
}
\caption{Total mass within $R_{200}$ against emission weighted temperature
  within 0.3 $R_{200}$ for simulated relations for dark matter
  concentration varying according to Equation \ref{beq:concentration}
  (solid), Equation \ref{beq:concentration2} (dashed) and a constant c=10
  (dotted). The excess energy for all the relations is zero. Observed data
  (crosses) are taken from the sample of \protect\citet{lloyd-davies00b}.}
\label{bfig:plot9}
\end{minipage}
\end{figure*}

Figure \ref{bfig:plot9} shows the data with overlayed simulated mass
temperature relations for different dark matter concentration variation
prescriptions. The solid line shows the simulated relation with the
concentration parameter varying with system mass as specified by Equation
\ref{beq:concentration}. It can be seen that this relation is much steeper
than the self-similar $T^{\frac{3}{2}}$ relation (dotted line), with a
logarithmic slope of $\sim 2$. The relation appears to be a reasonable fit
to the observed data, although if anything it is slightly steeper. The
dashed line shows the simulated relation with the concentration parameter
varying with system mass as predicted by \citet{navarro97a} for a
$\Lambda$CDM cosmology (see Equation \ref{beq:concentration2}). This
relation is flatter than that predicted by Equation
\ref{beq:concentration}, with a logarithmic slope of $\sim 1.75$, somewhat
flatter than is generally observed
\citep{sato00a,nevalainen00a,lloyd-davies00b}. It therefore seems that
observations generally favour concentration variation that is steeper than
the $\Lambda$CDM prediction. For this reason the observed relation Equation
\ref{beq:concentration} will used to derive the following the results but
comparisons with the $\Lambda$CDM prediction will be included where
necessary. The mass-temperature relation appears to be significantly
affected by the variation in concentration of galaxy clusters and these
effects should propagate into other relations involving galaxy cluster
temperatures.

\subsection{Beta-temperature relation}
\begin{figure*}
\begin{minipage}{140mm}
\centering{
\vbox{\psfig{figure=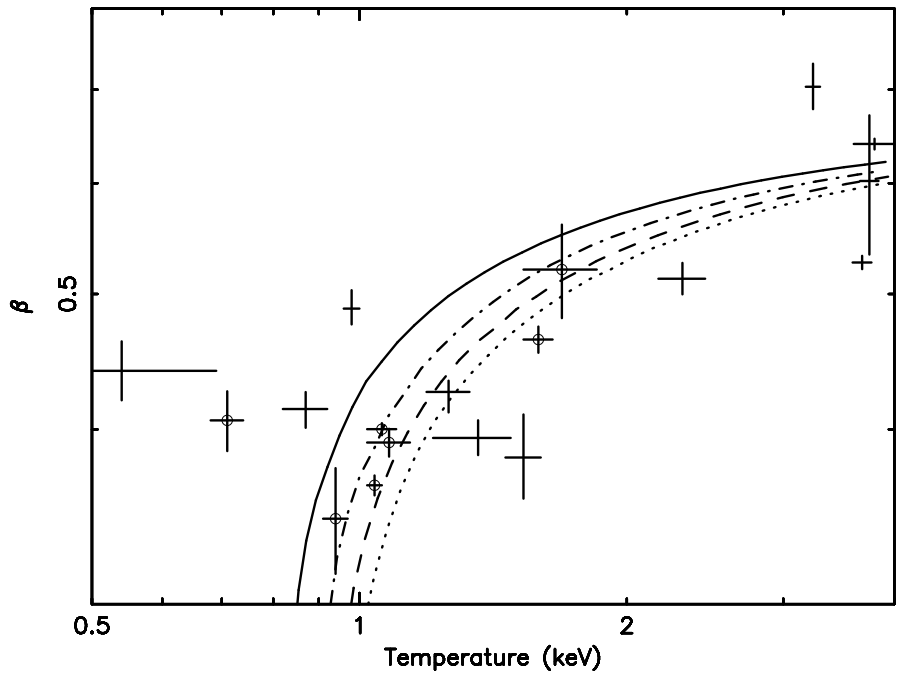}}\par
}
\caption{Asymptotic slope of the gas density profile, $\beta$, plotted 
  against emission weighted temperature within 0.3 $R_{200}$ for simulated
  relations with various dark matter concentrations; c=10 (solid), c=15
  (dot-dashed), c=20 (dashed) and c=25 (dotted). Observed data are taken
  from the samples of \protect\citet{helsdon00a} (circles) and
  \protect\citet{lloyd-davies00b} (crosses).}
\label{bfig:plot14}
\end{minipage}
\end{figure*}

The relationship between the asymptotic slope of gas density profile and
the temperature of galaxy clusters is important since it appears that the
main result of raising the specific energy of the ICM is to flatten the gas
density profile \citep{ponman98a,lloyd-davies00b}. Since we have previously
seen that variation in the dark matter concentration of galaxy clusters can
affect their emission-weighted temperatures, some effect on the
beta-temperature relation is also to be expected. Figure \ref{bfig:plot14}
shows $\beta$ plotted against emission weighted temperature within 0.3
$R_{200}$ for various dark matter concentration parameters.  The model
predictions are shown for concentration parameters of c=10 (solid), c=15
(dot-dashed), c=20 (dashed) and c=25 (dotted). In each case the specific
energy of the ICM has been raised by 0.75 keV per particle. The observed
data are taken from the samples of \citet{helsdon00a} (circles) and
\citet{lloyd-davies00b} (crosses). It can be seen that as the concentration
is increased, the relation is pushed to higher temperatures. However the
change in the relation is not particularly large, and cannot entirely
explain the scatter in $\beta-T$ observed.

The other parameter that can be varied is the amount by which the specific
energy of the ICM is raised. Figure \ref{bfig:plot13} shows $\beta$ plotted
against emission weighted temperature within 0.3 $R_{200}$ with the dark
matter concentration varying according to Equation \ref{beq:concentration}.
The model predictions for the specific gas energy raised by 0.25 keV per
particle (solid), 0.5 keV per particle (dot-dashed), 0.75 keV per particle
(dashed) and 1.0 keV per particle (dotted) are shown. It can be seen that
there appears to be a considerable amount of scatter with none of the
energy injection values chosen providing a good fit to all the data.  In
fact the statistical errors on the points accounting for only 5 percent of
the scatter about a mean fitted relation.  This result is not significantly
changed by removing the \citet{helsdon00a} points. If this scatter is
interpreted as due to a scatter in the amount of energy injected into each
system then energy injection covering at the range 0.25 and 1.0 keV per
particle are required to match the data. Only a small component of the
scatter can be attributed to scatter in the concentration parameters of the
systems.

\begin{figure*}
\begin{minipage}{140mm}
\centering{
\vbox{\psfig{figure=plot13.ps}}\par
}
\caption{Asymptotic slope of the gas density profile, $\beta$, plotted 
  against emission weighted temperature within 0.3 $R_{200}$ for simulated
  relations with specific energies raised to 0.25 keV per particle
  (solid), 0.5 keV per particle (dot-dashed), 0.75 keV per particle
  (dashed) and 1.0 keV per particle (dotted).  In all cases the dark matter
  concentration varies with mass according to the relation given in
  Equation \ref{beq:concentration}. Observed data are taken from the
  samples of \protect\citet{helsdon00a} (circles) and
  \protect\citet{lloyd-davies00b} (crosses).}
\label{bfig:plot13}
\end{minipage}
\end{figure*}

It is possible to use the predictions of the model to derive excess
energies for individual low mass systems.  Since the dark halo
concentration parameters of the systems affect the model predictions to
some extent, these measurements are best done for systems with known dark
matter concentrations. The sample of \citet{lloyd-davies00b} is the only
data available where concentration parameters have been measured for low
mass systems and also has the advantage of the various observational
parameters being derived within a consistent set of radii.  Excess energies
were therefore derived from the $\beta$ parameters and temperatures of the
eight systems in the sample of \citet{lloyd-davies00b} with temperatures
below 3 keV and are shown in Figure \ref{bfig:plot18} plotted against the
systems emission weighted temperatures within 0.3 $R_{200}$. It can be seen
that apart from the three lowest temperature systems the there is little
trend in excess energy with temperature although there is considerable
scatter. The mean excess energy is $\sim$1 keV per particle excluding the
three lowest temperature points. It is clear that the three lowest
temperature systems fall considerably below this value and show a strong
trend of decreasing excess energy with temperature. This trend is also
noticeable in Figure \ref{bfig:plot13}. While it is possible that this is a
real effect, the role of selection effect in this trend should not be
discounted.

\begin{figure*}
\begin{minipage}{140mm}
\centering{
\vbox{\psfig{figure=plot18.ps}}\par
}
\caption{Excess gas energy within $R_{200}$ plotted against emission 
  weighted temperature within 0.3 $R_{200}$ for the eight systems in the
  sample of \protect\citet{lloyd-davies00b} with temperatures below 
  3 keV. The dashed line is the model prediction for a constant luminosity
  of $2 \times 10^{41}$ erg s$^{-1}$ (see text).}
\label{bfig:plot18}
\end{minipage}
\end{figure*}

In particular the luminosity of a system decreases as excess energy
increases and this effect is much more pronounced in low temperature
systems.  For this reason if a sample is flux limited and there is some
distribution in excess energies, then there will be a tendency for systems
with low excess energies to be preferentially selected. The magnitude of
this effect will increase with decreasing system temperature since the
range of luminosities for a given range of excess energies is larger for
lower mass systems. The three lowest temperature systems may therefore
represent only the extreme low end of the distribution in excess energies
since low temperature systems with higher excess energies would not have
high enough luminosities to be part of the sample of
\citet{lloyd-davies00b}. It should be noted that one of the five higher
temperature systems has an excess energy higher than the mean by an amount
similar to the amount that the three low temperature systems are lower than
the mean. This suggests that there may be enough intrinsic scatter in the
excess energies of the systems to account for this effect.

We can approximate where this flux limit will lie in Figure
\ref{bfig:plot18} assuming that all these low mass systems are at
approximately the same redshift and there is some limiting luminosity below
which they cannot be observed. \citet{helsdon00a} provide an extensive
sample of X-ray bright galaxy groups all of which are brighter than $2
\times 10^{41}$ erg s$^{-1}$. Using out model a line of constant luminosity
in the excess energy-temperature plane can be calculated by adjusting the
excess energy until that luminosity is achieved for a range of system
masses. This luminosity limit of $2 \times 10^{41}$ erg s$^{-1}$ is shown
in Figure \ref{bfig:plot18} as the dashed line. It can be seen that in
general in the sample fall below the line where there luminosities should
be larger than $2 \times 10^{41}$ erg s$^{-1}$. The drop in excess energy
for the lowest mass systems also appears to shadow the luminosity limit in
the expected way. It therefore seems quite likely that this is a selection
effect rather than a intrinsic property of low mass galaxy groups.

\subsection{Luminosity-temperature relation}
A well studied relation between galaxy cluster properties is the
luminosity-temperature relation. This is primarily because the luminosity
and temperature are the two most easily measured properties of galaxy
clusters.  For the highest redshift clusters, which are extremely important
from a cosmological perspective, these are the only X-ray properties that
can be measured. It has long been know that the luminosity-temperature
relation does not follow the predictions of self-similar theory. For
self-similar clusters, luminosity would be expected to be proportional to
temperature squared \citep{kaiser86a}. However observational studies of the
luminosity-temperature relation tend to find steeper slopes, with
luminosity typically proportional to temperature cubed. For instance
\citet{edge91a} have measure the luminosity-temperature relation for a
sample of galaxy clusters to be $L \propto T^{2.62\pm0.10}$;
\citet{david93a} found $T \propto L^{0.297\pm0.004}$ and \citet{white97a}
found $T \propto L^{0.30\pm0.05}$.  These results are all in the region of
$L \propto T^3$. It has been suggested that this steepening from the
predicted relation might be caused by cooling flows \citep{allen98a}.
However \citet{markevitch98a} has derived a luminosity-temperature relation
corrected for the effects of cooling flows and measured a slope of
$2.63\pm0.27$. \citet{arnaud99a} measured the slope of the
luminosity-temperature relation for a sample galaxy clusters without strong
cooling flows to be $2.88\pm0.15$ and \citet{ettori00a} obtain a slope of
2.7 in the luminosity-temperature relation for a corrected sample of galaxy
clusters. It therefore seems clear that cooling flows cannot provide an
explanation for the steepening of the luminosity-temperature relation in
galaxy clusters.

Various authors have extended this work down to lower mass systems, galaxy
groups, and found that the luminosity-temperature relation steepens in
these systems. \citet{ponman96a} measured the slope of the
luminosity-temperature relation for Hickson compact groups to be
$8.2\pm2.7$ while \citet{helsdon00a} found the slope for a sample of loose
groups to be $4.9\pm0.8$. \citet{helsdon00b} have combined a sample of
loose and compact groups to derive a slope of $4.3\pm0.5$. This is
significantly steeper than the luminosity-temperature relation for rich
galaxy clusters.  This steepening of the luminosity-temperature relation in
galaxy groups has generally been attributed to the effect of heating by
galaxy winds \citep{cavaliere97a,helsdon00a,bower00a}. There appear to be
two separate effects on the luminosity-temperature relation which may or
may not be related. The steeper than predicted slope for high mass systems
and the steepening of the relation in low mass systems. It is therefore
extremely interesting to ask how the luminosity-temperature relation
predicted by our model compares with observations.

However in order to calculate the luminosity, some prescription is needed
for the emissivity of a plasma of a given temperature, density and
metallicity. We use the method of \citet{knight97a}, which uses bilinear
interpolation over the tabulated cooling function of \citet{raymond76a}, to
calculate the cluster luminosities and to derive the emission weighting for
the emission weighted temperatures.

\begin{figure*}
\begin{minipage}{140mm}
\centering{
\vbox{\psfig{figure=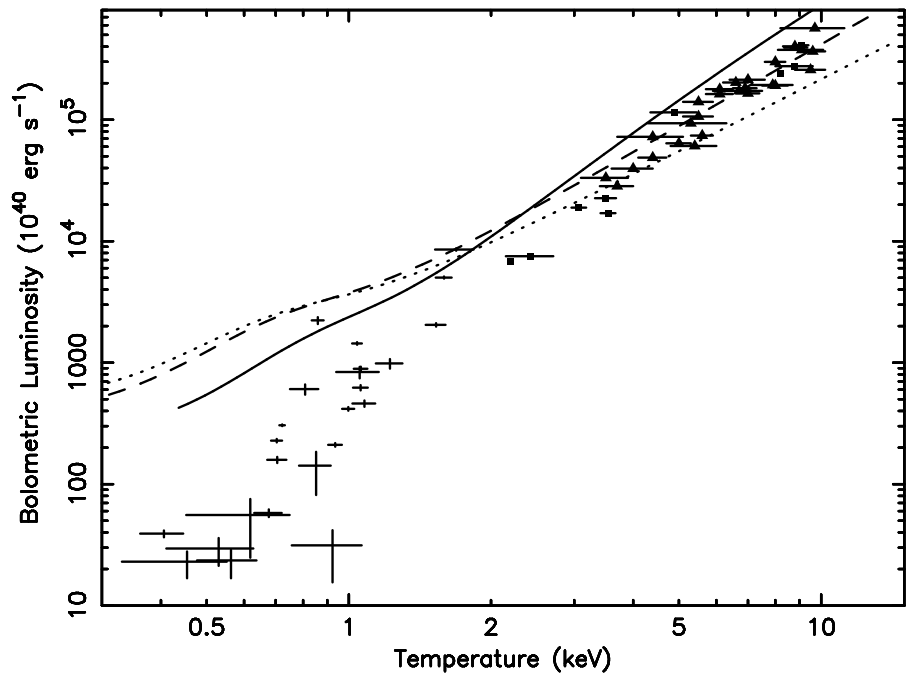}}\par
}
\caption{Bolometric luminosity within 0.3 $R_{200}$ against emission weighted 
  temperature within 0.3 $R_{200}$ for simulated relations for dark matter
  concentration varying according to Equation \ref{beq:concentration}
  (solid), Equation \ref{beq:concentration2} (dashed) and a constant c=10
  (dotted). Observed cooling corrected data are shown for the sample of
  \protect\citet{markevitch98a} (triangles). The sample of
  \protect\citet{arnaud99a} (squares) is uncorrected but was chosen not to
  contain strong cooling flows. The data shown as crosses is for the sample
  of \protect\citet{helsdon00a} and is not cooling corrected.}
\label{bfig:plot11}
\end{minipage}
\end{figure*}

\begin{figure*}
\begin{minipage}{140mm}
\centering{
\vbox{\psfig{figure=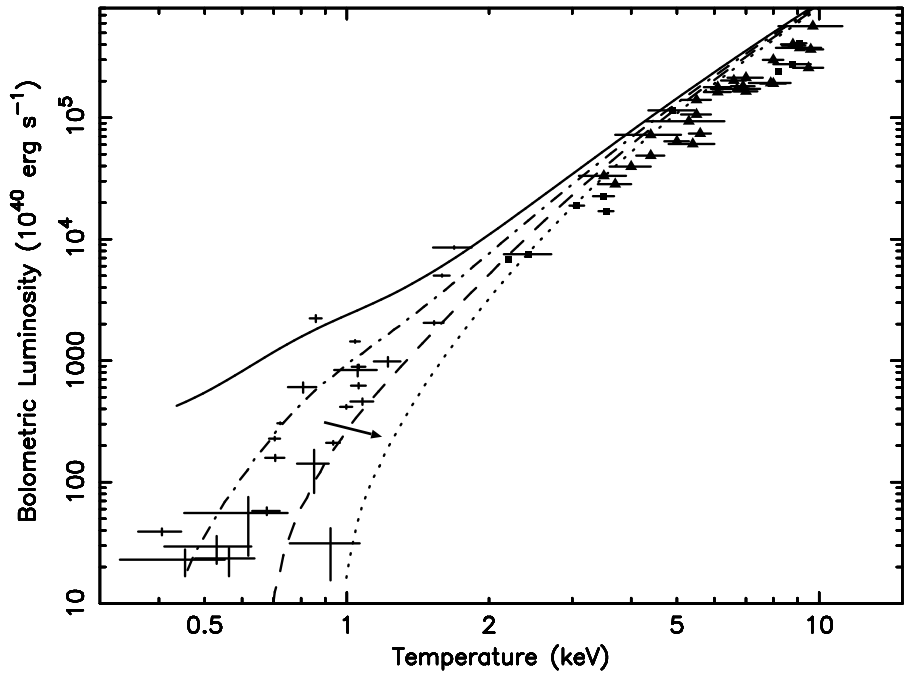}}\par
}
\caption{Bolometric luminosity within 0.3 $R_{200}$ against emission
  weighted temperature within 0.3 $R_{200}$ for simulated relations with
  excess gas energy of 0.0 keV per particle (solid); 0.25 keV per particle
  (dot-dashed) 0.5 keV per particle (dashed) and 0.75 keV per particle
  (dotted). In all cases the dark matter concentration varies with mass
  according to the relation given in Equation \ref{beq:concentration}.
  Observed cooling corrected data are shown for the sample of
  \protect\citet{markevitch98a} (triangles). The sample of
  \protect\citet{arnaud99a} (squares) is uncorrected but was chosen not to
  contain strong cooling flows. The data shown as crosses is for the sample
  of \protect\citet{helsdon00a} and is not cooling corrected. The arrow
  shows is an estimate of the average effect on the
  \protect\citet{helsdon00a} data if it was corrected for cooling.}
\label{bfig:plot12}
\end{minipage}
\end{figure*}

\begin{figure*}
\begin{minipage}{140mm}
\centering{
\vbox{\psfig{figure=plot16.ps}}\par
}
\caption{Asymptotic slope of the gas density profile, $\beta$, plotted 
  against emission weighted temperature within 0.3 $R_{200}$ for dark
  matter concentration varying according to the relation given in Equation
  \ref{beq:concentration} and raised specific energy of 0.75 keV per
  particle. The solid line is the relation for the temperature boundary
  condition of $0.5 T_{vir}$. The dashed line is the relation when a
  minimum temperature of $\frac{2}{3}\Delta{E}$ is imposed on the boundary
  condition.  Observed data are taken from the samples of
  \protect\citet{helsdon00a} (circles) and \protect\citet{lloyd-davies00b}
  (crosses).}
\label{bfig:plot16}
\end{minipage}
\end{figure*}

Figure \ref{bfig:plot11} shows the bolometric luminosity within 0.3
$R_{200}$ against emission weighted temperature within 0.3 $R_{200}$.  The
dotted line shows the model prediction for systems with a fixed dark matter
concentration of c=10 and no heating of the ICM. This is compared with
observed data for galaxy clusters from \citet{markevitch98a} (triangles)
and \citet{arnaud99a} (squares). The data of \citet{markevitch98a} is
corrected for cooling and that of \citet{arnaud99a} is selected to contain
no systems significantly affected by cooling. Observed data for galaxy
groups is taken from the sample of \citet{helsdon00a} (crosses) and is not
corrected for cooling. It can be seen that the model relation is
considerably flatter than the observed data. To investigate the effect of
varying the concentration parameter on the luminosity-temperature relation,
the relations for dark matter concentration varying according to Equations
\ref{beq:concentration2} and \ref{beq:concentration} are shown as the
dashed and solid lines respectively. The effect of varying the
concentration is to significantly steepen the luminosity-temperature
relation, bringing it into much better agreement with the observed data.
The relation with concentration varying according to Equation
\ref{beq:concentration2} has a logarithmic slope of $\sim 2.2$ above 2 keV,
the one varying according to Equation \ref{beq:concentration} has a
logarithmic slope of $\sim 2.7$ above 2 keV.  It can be seen that there is
some variation in the slope with system temperature. The relation with a
constant concentration parameter of c=10 (dashed line) has a logarithmic
slope of $\sim 1.97$ although again there is some variation in the slope
with system temperature. It should be noted that strictly the expectation
that the luminosity-temperature relation will have a logarithmic slope of 2
is only true if the emission process is pure thermal bremsstrahlung. The
contribution of emission lines at low temperatures will tend to flatten the
relation so that the relation will be expected only to asymptote to a
logarithmic slope of 2 at high temperatures where the effect of emission
lines is less important.

The normalisation of the luminosity-temperature relation for varying dark
matter concentration in Figure \ref{bfig:plot11} appears somewhat higher
than the mean trend of the observed data. One possible reason for this is
that the relation we use for the core radius in the model is taken from
\citet{lloyd-davies00b}, whose sample was picked to be as relaxed as
possible. It is possible that the value of 7 percent of $R_{200}$ which we
use is not representative of galaxy cluster population as a whole and that
less relaxed clusters have significantly larger core radii. Previous
studies that did not limit themselves to relaxed systems, have measured
core radii that are considerably larger \citep{mohr99a}. Increasing the core
radius for a fixed gas fraction will reduce the central gas density and
therefore the luminosity. Another related point is that the canonical value
of $\beta=\frac{2}{3}$ that we use as the default in the model when no
heating has occurred does not have a great deal of theoretical support.  If
in fact the mean value of $\beta$ in real high mass clusters never quite
reaches the canonical value and this would also reduce the luminosity for a
given gas mass. It should be noted that the mean $\beta$ for the systems
above 4 keV in the sample of \citet{lloyd-davies00b} is 0.61. It is also
possible that the gas fraction of 0.2 that we are using is somewhat higher
than the mean value in the systems to which we are comparing the model.

It can be seen in Figure \ref{bfig:plot11} that while the variation of the
concentration parameter results in the high temperature part of the
relation being a reasonable fit to the data the low temperature part of the
relation still over-predicts the luminosities of the low temperature
systems. To investigate whether ICM heating can improve the fit for the low
temperature systems relations were plotted with varying amounts of energy
injection into the ICM. Figure \ref{bfig:plot12} shows the
luminosity-temperature relations for energy injection of 0.0 keV per
particle (solid); 0.25 keV per particle (dot-dashed) 0.5 keV per particle
(dashed) and 0.75 keV per particle (dotted). It can be seen that the
heating the ICM by up to 0.75 keV per particle has little effect on the
luminosity-temperature relation for the galaxy clusters. However the
heating does have a significant effect on the luminosity-temperature
relation for the galaxy groups. As is to be expected the flattening of
the gas density profile reduces the luminosity of the low mass systems
towards the observed relation. It appears that energy injection of between
0.25 and 0.5 keV per particle best reproduces the observed relation.
It should be noted that there is considerable scatter in the observed
luminosity-temperature relation in galaxy groups, but given the possibility
of scatter in both the injected energy and concentration parameter this
is not surprising.

One discrepancy in this analysis is that the galaxy group data is affected
to some extent by cooling, while the model does not take cooling into
account. The luminosity-temperature relation is particularly susceptible to
the effects of cooling since both parameters will be affected by it.  We
have attempted to to quantify the effects of this on our results using the
observed properties of cooling flows in groups.  \citet{helsdon01a} find
that these contribute typically 25 percent of the group luminosity, whilst
\citet{lloyd-davies00b} find that removing the effects of cooling from
spatial-spectral models fitting group X-ray data results in an increase in
mean temperature of about 20 percent.  The estimated total average effect
of correcting the galaxy group data for cooling is shown as an arrow in
Figure \ref{bfig:plot12}. It can be seen that with this correction, most of
the data would be consistent with a mean injection energy of between 0.5
and 0.75 keV per particle.

\subsection{Effects of boundary conditions}
\label{sec:boundary}
The results presented so far are for a specific set of plausible boundary
conditions, given our present knowledge. The temperature at $R_{200}$ is
fixed at $0.5 T_{vir}$ where $T_{vir}$ is defined by Equation \ref{eq:tvir}
and the density at $R_{200}$ is constant. However it is possible that these
boundary conditions are over simplified. To test the effect of these
assumptions the boundary conditions must be varied in order to see how much
difference they make to the results. The temperature boundary condition of
$0.5 T_{vir}$ in particular is based only on numerical simulations rather
than an empirical evidence. Not only is it possible that 0.5 is the wrong
normalization but more importantly it is possible that it may not scale
with $T_{vir}$.  For example, if the intergalactic medium (IGM) has been
heated before the cluster forms, in the absence of significant cooling the
minimum temperature at $R_{200}$ will be the temperature to which the IGM
was heated. In reality this represents an upper limit to the temperature
constraint, since if the IGM was heated a long time before the system
collapsed, the Hubble expansion would subsequently lower the gas
temperature.  In low mass systems this constraint would result in the
temperature at $R_{200}$ having a lower limit of $\frac{2}{3}{\Delta}E$
where ${\Delta}E$ in the energy injection in keV per particle.  To assess
how altering to the temperature boundary condition might affect the
results, $\beta$-temperature relations with and without a minimum
temperature were simulated for an energy injection of 0.75 keV per particle
and dark matter concentration varying according to Equation
\ref{beq:concentration}.  This is shown in Figure \ref{bfig:plot16}. The
solid line shows the relation with no minimum temperature and the dashed
line shows the effect of imposing a minimum.

It can be seen that the result of introducing a minimum temperature to the
temperature boundary condition is to reduce the change in $\beta$ that is
needed, to get a specified excess energy, for systems below a certain
emission-weighted temperature. In this case at $\sim 2$ keV the
temperature at $R_{200}$ hits the limit of $\frac{2}{3}\Delta{E}$ and
systems below this emission-weighted temperature have higher gas
temperatures at $R_{200}$ in the raised specific energy case than the
default case. This causes more energy to go into heating the ICM and less
into flattening the gas density profile resulting in higher values of
$\beta$. Increasing the minimum temperature at the outer boundary of the
model will increase the emission-weighted temperature below which the
$\beta$ parameter deviates from the predictions of our standard model.
The mechanism by which the effect of energy injection on the
boundary condition has been implemented is very crude, but it does give
good indication of the general effect that this sort of process would have.
The effect is to cause the decrease in $\beta$ with decreasing temperature
to become less drastic below a certain temperature. While the number of
very low temperature systems in our sample is not that large, the data do
suggest a steep decrease in $\beta$ which flattens off below $\sim 1$ keV.
However as noted previously there are likely to be strong selection effects
at work in the lowest mass system which might be expected to cause only the
low mass systems with the highest $\beta$ parameters to be observable.

\section{Discussion}
We have modeled two major processes that result in the breaking of the
expected self-similarity of galaxy clusters; energy injection and variation
in the dark matter halo concentration. Both these processes are shown to
have significant effects on the properties of galaxy clusters. The mass
temperature relation is not significantly affected by heating of the ICM
except for the lowest mass systems whose temperatures are increased.
However variation of the dark matter concentration significantly alters the
normalisation of the mass-temperature relation.  The observed trend in
concentration parameter with mass \citep{lloyd-davies00b}, results in a
steepening of the mass-temperature relation from its expected logarithmic
slope of 1.5 to $\sim 2$. This is a good match the observed data, where
slopes of this order are observed: $1.96\pm0.21$ \citep{lloyd-davies00b},
$1.79\pm0.14$ \citep{nevalainen00a} and $2.04\pm0.04$ \citep{sato00a}. The
observed mass-temperature relation therefore appears to be well matched by
the model predictions when systematic variation in dark matter
concentration with system mass is included.

The relation between the asymptotic slope of the gas density profile and
the emission weighted temperature has generally been interpreted as due to
energy injection into the ICM. The model produces relations that roughly
follow the trend in the data for energy injection in the range 0.5 to 1.0
keV per particle. However the relation has a considerable amount of scatter
and no one model appears to be consistent with all the data. This may be
due to real scatter in the amount of excess energy injected into the
systems, although it is possible that the errors on the observational data
have been underestimated. The effect of varying the dark matter
concentration is to push the relation to higher temperature.  The inclusion
of systematic variation of the concentration parameter with system mass as
observed by \citep{lloyd-davies00b} has the effect of slightly lowering the
amount of energy that in needed to get a given $\beta$ value. Scatter in
the concentration of dark matter halos may make some contribution to the
scatter in the beta-temperature relation. However it does not appear to be
a large enough effect to be the dominant cause.

The luminosity-temperature relation of the model has the expected
logarithmic slope of $\sim 2$ when the model is self-similar. However
introducing the systematic variation in dark matter concentration observed
by \citet{lloyd-davies00b} steepens the slope to 2.7 which is comparable
with the slopes of 2.6 to 2.9 observed by recent authors
\citep{markevitch98a,arnaud99a,ettori00a}. This is to be expected as
increasing the concentration of the low mass systems will increase their
emission weighted temperatures and so steepen the relation. It therefore
seems clear that the steepening of the slope of the luminosity-temperature
relation in the cluster regime is largely, if not entirely, explained by
the systematic variation of dark matter halo concentration with system
mass. It should also be noted that the systematic trend seen by
\citet{lloyd-davies00b} has considerable scatter which is expected
theoretically as low mass systems are expected to form a higher redshift
only on average. Scatter in the concentration of dark matter halos could
contribute a considerable amount to the scatter in the
luminosity-temperature relation. However cooling flows are also likely to
contribute a considerably component to the scatter.

Models including systematic variation of the dark matter concentration with
system mass reproduces the slope of the luminosity-temperature relation for
galaxy clusters but does not predict the steepening of the relation seen in
galaxy groups. However the addition of heating of the ICM steepens the
relation in galaxy groups as flattening the gas density profile reduces the
luminosity and increases the emission weighted temperature. There is
considerable scatter in the observed relation but most systems appear to
fall in between the model relations for energy injection of 0.5 keV and
0.75 keV per particle after correcting for the effects of cooling in
groups. It should be noted that for galaxy groups, scatter in both dark
matter concentration and energy injection may contribute to the scatter in
the relation along with cooling flows and other factors.

It therefore appears that the mean trends in the mass, $\beta$ and
luminosity with system temperature can be explained reasonably well by a
combination of variation in the dark matter halo concentration and energy
injection into the ICM. The energy injection that best reproduces these
trends is in the range $\sim 0.5-0.75$ keV per particle. This is comparable
with the value of 0.44 keV per particle found by \citet{lloyd-davies00a}
using analytical models fitted to X-ray spectral images of galaxy clusters
and groups.  \citet{bower00a} derive a value of 0.6-1 keV per particle when
comparing their model with observational data which is also comparable with
our result. There is considerable scatter in all the relations which may be
due to a combination of scatter in the concentration parameter and energy
injection.

In the case of the $\beta$-temperature relation it is possible to derive
excess energies for individual systems and in general these are higher than
the values derived for the relations as a whole. The lowest mass systems
however appear to have systematically lower excess energies and it is
possible that this is due to selection effects where low mass systems with
high excess energies are not observed because they are not luminous enough.
The mean excess energy of the systems below 3 keV excluding the three
lowest mass systems is $\sim$1 keV per particle. This suggests the
possibility that care must be taken when deriving mean excess energies from
cluster relations, since the lowest mass systems which tend to have the
largest influence on the result may have systematically lower excess
energies due to selection effects. A value of $\sim 0.5$ keV per particle
therefore appears to be a lower limit on the mean excess energy in galaxy
clusters and groups, and the true mean value may be as high as 1 keV per
particle.

It is possible to estimate the amount of energy that could be available as
a result of star formation in order to assess whether it is possible for
this process alone to account for the excess energy we observe. Assuming
that each supernova produces $10^{51}$ erg of kinetic energy
\citep{woosley86a}, there is 0.007 supernova per $M_{\odot}$ of stars
\citep{bower00a} and the gas and stellar fraction of galaxy clusters and
groups are 0.2 and 0.11 respectively \citep{lloyd-davies00a}, results in an
energy per particle of 1.2 keV. There are considerable uncertainties in the
assumptions made in this derivation but it does suggest that it is possible
for star formation alone to provide the excess energy we observe, although
the higher mean excess energy of $\sim$1 keV per particle we derive for
individual systems would require a very high efficiency in transferring
the energy from supernovae to the intergalactic medium. Detailed observation
of the metal enrichment of the ICM \citep{finoguenov00a} also support a
scenario of heating of the ICM prior to cluster formation by galaxy winds.

It seems clear that the approach of modeling similarity-breaking in galaxy
clusters and comparing the predictions to observations is extremely useful
in constraining the processes involved in the formation and evolution of
galaxy clusters. With the advent of observation of large samples of galaxy
clusters using XMM-Newton and Chandra it should be possible to construct
much more detailed models which place much greater constraints on the
structure of galaxy clusters. It may also be possible to directly test the
temperature profiles predicted by our models with such observations.

\section*{Acknowledgments}
The authors would like to thank Stephen Helsdon for providing electronic
versions of his published data. This work made use of the Starlink
facilities at Birmingham and Durham.  EJLD acknowledges the receipt of a
PPARC studentship.

\bibliography{beta}

\label{lastpage}

\end{document}